\documentstyle[aps,preprint]{revtex}
\tighten 
\begin{document}
\draft
\preprint{DFUP-118/96, gr-qc/9612030}

\title{Real and complex connections for canonical gravity} 
\author{Giorgio Immirzi\footnote{e--mail: immirzi@perugia.infn.it}}
\address{Dipartimento di Fisica, Universit\`a di Perugia, and
I.N.F.N., sezione di Perugia\\ via A. Pascoli, 6100 Perugia, Italy}
\date{December 2, 1996}
\maketitle 
\begin{abstract} 
Both real and
complex connections have been used for canonical gravity: the complex
connection has $SL(2,C)$ as gauge group, while the real connection has
$SU(2)$ as gauge group. We show that there is an arbitrary parameter
$\beta$ which enters in the definition of the real connection, in the
Poisson brackets, and therefore in the scale of the discrete spectra
one finds for areas and volumes in the corresponding quantum theory. A
value for $\beta$ could be could be singled out in the quantum theory
by the Hamiltonian constraint, or by the rotation to the complex
Ashtekar connection.
\end{abstract}
 
%
\section{Introduction}
 A lot of recent work in canonical quantum
gravity is based on using as canonical variables the pair
$(E^{ia},A^i_a)$, where $E^{ia}$ is a triad of density weight one and
$A^i_a$ a real  $SU(2)$ connection; these variables were introduced by
Barbero\cite{Ba} as an alternative to the complex Ashtekar
variables\cite{Ai}.  For $SU(2)$ to be the gauge group for general
relativity we have to fix the local inertial frames,  choosing what is
called "time gauge"; once this choice is made, it will be shown that
there is a certain freedom in the choice of the connection,
parametrized by a variable $\beta$. No particular real real value of
$\beta$ has a geometric motivation, and any one gives an acceptable,
if somewhat bizarre, $SU(2)$ connection. The choice $\beta =i$ gives
the Ashtekar variables, requires that we impose reality conditions on
the variables, but restores the full invariance of the canonical
theory under local Lorentz transformations; as an additional benefit,
this choice simplifies  the expression of the hamiltonian constraint
considerably.

The difference between these choices becomes important when we attempt
to quantize the theory. It is difficult to find a scalar product if
the gauge group is $SL(2,C)$, even in simplified cases\cite{GI}. On
the contrary, for gauge invariant functionals of a $SU(2)$ connection,
the metric being well defined, one can study the spectrum of operators
that can be interpreted as representing the area of a surface and the
volume of a region in space\cite{ARS}; both these spectra turn out to
be discrete\cite{DPR,AL,L,Tc}, indicating a discrete structure of
space at the Planck scale. Unfortunately, the arbitrary parameter
$\beta$ multiplies the scale of these spectra, and therefore the
results are meaningless unless we find a way to fix this constant.
Since the constant $\beta$ appears explicitly in the expression of the
Hamiltonian constraint, it may be that that solutions to this
constraint exist only for particular values, most likely $\beta =1$.
An alternative, but strictly related possibility would be that $\beta
=1$ is required because it is the value for which a rotation to the
Ashtekar variables is possible\cite{Aii,Ta}, restoring the gauge group
$SL(2,C)$. Both these goals have been attempted in recent work
\cite{Tb}.

\section{Connections.}
 A canonical formulation of general relativity
may be obtained starting from the action $S={1\over
4\kappa}\int\epsilon_{IJKL}e^I\wedge e^J\wedge R^{KL}$ (where
$I,J,\dots =0,1,2,3$, $e^I$ are vierbeins, $ R^{KL}$ is the curvature
2--form of the Levi--Civita connection $\Omega^{IJ}$, and  $\kappa
:=(8\pi G_{Newton})/c^3$), then taking space slices $\Sigma_t:\
t(x)=$const. with a vector $t^a: \ t^a\partial_at=1$  related to the
unit normal to the slice $n^a$ by $t^a=Nn^a+N^a$, or
$n_a=-N\partial_at$. If at this stage we partially fix the $O(3,1)$
gauge freedom choosing "time gauge", i.e. setting:
 \begin{equation}
   e^0_a=-n_a=N\partial_a t  \label{ci}
 \end{equation}
 we are left with
invariance under local $O(3)$ transformations with $\ e^{ia},\
i=1,2,3$  space--like  providing a local frame on the slice and the
inverse 3--metric $q^{ab}=e^{ia}e^{ib}$.   The pull--back of
$\Omega^{IJ}$ to the space slice gives the 3--d Levi--Civita
connection $\Gamma^i_a:={\textstyle 1\over 2} q_a^{\
b}\epsilon_{jik}\Omega^{jk}_b$, and the extrinsic curvature $K^i_a:=
q_a^{\ b}\Omega_b^{0i}=e^{ib}K_{ab}$.  Instead of the A.D.M. variables
$(q_{ab},\pi^{ab}:=\sqrt{q}(K^{ab}-q^{ab}K)) $, we  can take the
pair\cite{Aii}: 
\begin{equation}
 (E^{ia}:= {\textstyle 1\over
2}\tilde\epsilon^{abc}\epsilon_{ijk}e^j_be^k_c= \sqrt{q}\; e^{ia}\
,\quad  K^i_a) \label{cii} 
\end{equation}
 From the definitions it follows that for small variations: 
\begin{eqnarray} K^i_a\;\delta
E^{ia}&=&{1\over 2\sqrt{q}}K_{ab}\delta (E^{ia}E^{i})=\nonumber\\
&=&{1\over 2q}(\pi_{ab}-{\textstyle 1\over 2}\pi\; q_{ab})\delta
(qq^{ab})= -{\textstyle 1\over 2}\pi^{ab}\;\delta q_{ab}
\end{eqnarray} 
and therefore the only non vanishing Poisson-brackets are: 
\begin{equation}
\{K^i_a(x),E^{jb}(y)\}=\kappa\delta^b_a\delta^{ij}\delta(x,y)
\label{ciii} 
\end{equation}
 With the connection $\Gamma^i_a$ we can
compute covariant derivatives $D_a$ and the curvature  $R^i_{ab}$, and
write the constraints of the theory in the form:
 \begin{eqnarray}
{\cal G}_i&:=&\epsilon_{ijk}K^j_aE^{ka}\approx 0\nonumber\\ {\cal
V}_c&:=&q_{bc}\nabla_a\pi^{ab}=2E^a_i D_{[a}K^i_{c]}\approx 0
\label{civ} \\ {\cal
H}&:=&{1\over\sqrt{q}}\big(\pi^{ab}\pi_{ab}-{\textstyle 1\over 2}
\pi^2-q\;R\big) =\nonumber\\ &=& {1\over\sqrt{\det E}} \left(
2E^{[a}_iE^{b]}_jK^j_aK^i_b+ \epsilon_{ijk}E^{ia}E^{jb}R^i_{ab}
\right) \approx 0\nonumber 
\end{eqnarray} 
The first constraint has been added to the A.D.M. ones to make sure
 that $K_{ab}$  is symmetrical.

The trouble with this formulation is that $\Gamma^i_a$ is a derived
quantity, that can be expressed in terms of $E^{ia}$, its derivatives
and its inverse solving the 9 linear equations it satisfies:
\begin{equation}
 D_aE^{ia}=0\qquad ;\qquad
\epsilon_{ijk}E^{a}_iE^{(b}_jD_aE^{c)}_k=0  \label{cv}
 \end{equation}
To obtain a connection that is in some sense a conjugate quantity to
the triads we may follow\cite{Ba}  and change our basic variables, for
some $\beta$ to be fixed, to:
 \begin{equation}
  (E^{ia}\ ,\
A^{(\beta)i}_a:=\Gamma^i_a\,+\,\beta\, K^i_a ) \label{cvi}
\end{equation}
 By the properties of the Levi--Civita connection, we find:
 \begin{eqnarray} 
 E^{ia}\;\delta A^{(\beta)i}_a&=&\beta
E^{ia}\;\delta K^i_a+E^{ia}\;\delta\Gamma^i_a=\nonumber\\ &=& \beta
E^{ia}\;\delta K^i_a+\partial_a(\tilde\epsilon^{abc} e^k_b\;\delta
e^k_c ) \label{cvii} 
\end{eqnarray} 
so that the only non vanishing Poisson brackets are:
\begin{equation}
  \{ A^{(\beta)i}_a(x),E^{jb}(y)\} =\beta\kappa\,
\delta^i_j\delta^b_a\delta(x,y) \label{cviii} 
\end{equation} 
Unlike $K^i_a$,  $A^{(\beta)i}_a$ transforms like  an $SU(2)$ 
connection for any $\beta$, so that one can have $D^{(\beta )}_a$
derivatives,  a curvature $F^{(\beta )i}_{ab}$, and for any path
$\gamma$ a "transporters" $g_\gamma :=P\exp (\int \tau_iA^{(\beta )
 i}_a d\gamma^a)$. Geometrically, all this is perhaps a bit
artificial: we are using a connection with torsion, and we are 
smuggling the dynamics of the theory in this torsion.

With some algebra one finds that the constraints eq.~(\ref{civ}) can
be rewritten in the form: 
\begin{eqnarray}
 &&D^{(\beta
)}_aE^{ia}=\beta{\cal G}_i\  \approx\ 0  \nonumber\\ &&E^{ia}F^{(\beta
)i}_{ab}=\beta\;{\cal V}_b+\beta^2\;K^j_b{\cal G}_j\ \approx\ 0
\label{cix}\\ &&{\cal H}= {\epsilon_{ijk}\over\sqrt{\det
(E)}}E^{ia}E^{jb}F^{(\beta )k}_{ab} - 2{(1+\beta^2)\over\sqrt{\det
(E)}} E^{[a}_iE^{b]}_jK^i_aK^j_b \nonumber\\
&&\qquad\qquad\qquad\qquad\qquad\qquad\qquad +\ldots \approx 0
\nonumber 
\end{eqnarray}
 where the dots are terms proportional to
$\beta{\cal G}_i$ and its derivatives.

Clearly, complications will come from the last constraint, in
particular from its overall factor $1/\sqrt{\det (E)}$ and from the
messy term proportional to $(1+\beta^2)$. The quickest way to get rid
of it would be to take $\beta=i$, which is Ashtekar's original choice
\cite{Ai}(together with the idea of absorbing the factor $1/\sqrt{\det
(E)}$ in the Lagrange multiplier). But this choice can be motivated by
much more than the simplification of the constraints: in fact for this
value of $\beta$ the variables eq.(\ref{cvi})  can be defined quite
generally, {\it without} the gauge choice eq.~(\ref{ci}), as the
pull--back to the space slice of the self--dual connection, and of the
self--dual product of two vierbeins:
 \begin{eqnarray}
A^{(i)i}_a&:=&q_a^{\ b}C^i_{IJ}\Omega_b^{IJ}\quad :=q_a^{\ b}(
-{\textstyle{1\over 2}}\epsilon_{ijk}\Omega^{jk}_b+i\Omega^{0i}_b) \\
E^{ia}&:=& -\epsilon^{abc}C^i_{IJ}e^I_be^J_c =-2i\,{\det(e^I_a)\over
N}\, C^i_{IJ}\,n_b e^{Ib}e^{Ja} \label{cx} 
\end{eqnarray}
 here $C^i_{IJ}$ projects antisymmetric tensors to the $(1,0)$
representation of $SL(2,C)$, i.e. selects the self--dual part, so that
$\epsilon_{IJ}^{\ \ KL}C^i_{KL}=2iC^i_{IJ}$. This definition of the
basic variables coincides with eq.~(\ref{cii}) if we assume "time
gauge", eq.~(\ref{ci}), but we have to add to the constraints
eq.~(\ref{cix}) the reality conditions: 
\begin{equation}
E^{ia}E^{ib}={\rm\ real}\quad ;\quad
\epsilon_{ijk}E^{a}_iE^{(b}_j\,D_a^{(i)}E^{c)}_k={\rm\ imaginary}
\label{cxi} 
\end{equation}
 On the other hand, adopting this connection
we retain the Lorentz group as gauge group of the canonical
theory\footnote{It is not often emphasized that the derivation of the
Ashtekar variables does not require any gauge fixing, but it is
obvious from their original definitions; for ex. the alternative
definition given for $E^{ia}$ in eq.(\ref{cx}) is a direct
transcription of eq.1.13b  of \cite{Abib}}.

Because of the obvious complications of dealing with complex variables
and with the non-compact gauge group $SL(2,C )$,  the alternative
choice $\beta=1$, the "Barbero connection",  has been used in all the
more recent work. In Euclidean (++++) gravity the second term of $\cal
H$ is proportional to $(1-\beta^2)$, and the choice $\beta =1$ {\it
is} natural, just like $\beta =i$ was "natural" for the (--+++)
signature, but that is not particularly relevant, unless somehow we
learn to "Wick--rotate" the theory. Otherwise, there is no obvious
geometric meaning, and nothing special about the value $\beta=1$, and
and we can leave $\beta$ arbitrary (but real $>0$) in the following.

One can also introduce a Poisson bracket preserving map on the algebra
of functions on phase space which, in a certain sense, allow us to
change $\beta$. Following Thiemann\cite{Ta}, one  notices that
\begin{equation}
 T:=\int_\Sigma K^i_aE^{ia}d^3x 
 \end{equation} 
from eq.~(\ref{ciii}) and the properties of Levi--Civita connections
has the following Poisson brackets: 
\begin{equation} 
\{T,E^{ia}\}=E^{ia}\
;\quad \{T,\Gamma^i_a\}=0\ ;\quad\{T,K^i_a\}=- K^i_a   \label{cxii}
\end{equation} 
It follows that if we indicate the Hamiltonian evolution induced
by some $H$ on a function $f$ on phase space  by the map: 
\begin{equation} W_H(t) \circ f:=f+t\{f,H\} +{t^2\over
2!}\{\{f,H\},H\}+\ldots \label{cxiii} 
\end{equation}
 we shall have:
\begin{equation}
 W_T(t)\circ E^{ia}=e^{-t}E^{ia}\ ;\quad W_T(t)\circ
A^{(\beta)i}_a=\Gamma^i_a+\beta e^tK^i_a  \label{cxiv} 
\end{equation}
At a more speculative level,  we can think of choosing $t=i\pi /2$ and
using this map to perform a rotation from the $\beta=1$ to the
Ashtekar connection \cite{Ta,Aii}, similar to a "Wick rotation", a
possibility that is particularly exciting in the quantum case.
\section{Quantization.}
 To quantize the theory we may use the
connection representation, in which states are functionals of
$A^{(\beta )i}_a(x)$, and 
\begin{equation}
 E^{ia}\quad\to\quad \hat
E^{ia} := {\beta\kappa\hbar\over i}{\delta\over \delta A^{(\beta
)i}_a} \label{si} 
\end{equation} 
The important states turn out to be
"spin net states", which are an obvious generalization of Wilson
loops. Given a net $n$ in 3--space with $V$ vertices joined by
(analytic) paths $\gamma_1,\ldots,\gamma_L$, we assign: to each line
an orientation, hence a "transporter" $g_l$, and a spin
$s_l=0,{\textstyle{1\over 2}} ,1,\ldots$, so that $\gamma_l\to{\cal
D}^{s_l}_{mm'}(g_l)$; to each vertex $v$ an $SU(2)$ invariant tensor
$C^v_{m_a,..,m_d}$, in such a way that: 
\begin{equation}
\psi_{\underline n}(\{g_l\})=\sum_{ \{ m \} }\prod_v C^v_{...}\prod_l
{\cal D}^{s_l}_{..}(g_l)    \label{sii} 
\end{equation} 
is gauge invariant.
 If $g_l\in SU(2)$, and we indicate by $dg_l$ the Haar
measure, the $\psi_{\underline n}$ that we can associate to a given
net form a Hilbert space with the scalar product: 
\begin{equation}
<\psi_{{\underline n}'}|\psi_{\underline n} >
=\int{\overline\psi}_{{\underline n}'}\;\psi_{\underline n} \prod dg_l
\label{siii} 
\end{equation}
 Ashtekar and collaborators have discovered
(see e.g.\cite{AL}) that if one considers all possible nets, then the
set of these states is {\it dense} in the Hilbert space of gauge
invariant functionals of  $A^{(\beta )}$. In this sense
eq.(\ref{siii}) induces a measure $DA^{(\beta)}$ in this space. One
can therefore define a Rovelli--Smolin "loop transform"\cite{RSa}, and
focus on states which have support on "weaves"\cite{ARS}, huge nets
with mesh sizes of the order of the Planck length. In recent work
these states have been used to find the spectrum of the operators that
correspond to the area of a surface $S$ and to the volume of a  region
$R$, and to investigate the operator form of the constraints.

Very briefly: if the surface $S$ intersects a subset  $\cal L$ of
lines of the net, does not touch the vertices, has no line lying on
it, carefully regularizing the operator ({\it first} taking the square
root, {\it then} removing the regulator), one finds :
 \begin{eqnarray}
\hat A (S)\psi_{\underline n}\ &=&\ :\int_sd^2\sigma \sqrt{n_an_b\hat
E^{ia} \hat E^{ib} }\; :\;\psi_{\underline n}\ =\nonumber\\\ &=&
(\beta \hbar\kappa )\sum_{l\in {\cal
L}}\sqrt{s_l(s_l+1)}\;\psi_{\underline n} \label{siv} 
\end{eqnarray}
This basic result has been extended in various directions, in
particular to the volume operator\cite{DPR,AL,L,Tc}. On the other
hand, there is no known way to extend it to the theory based on the
Ashtekar connection: all operators seem to have the wrong hermiticity,
whatever scalar product we use.

However, it is clear from eq.~(\ref{siv}) that the discrete spectra
one gets for areas and volumes \underbar{cannot} at this stage be
interpreted as evidence for a discrete structure of space, because of
the arbitrariness of $\beta$. The key ingredients we have used are the
gauge invariance and the commutation relations that correspond to the
Poisson brackets eq.~(\ref{cviii}), and  we must conclude that by
themselves they are not enough to fix the scale of the theory.

The quantization program requires therefore that we find some good
reason to fix $\beta$. I can imagine two ways in which this could
happen. One could argue that a larger group than $SU(2)$ is necessary,
as suggested for instance by the experience of current algebra, and
that one should insist on using the Ashtekar connection, overcoming
the dificulties that have been found; or that the states considered
should be restricted by more than just gauge invariance, in particular
by the Hamiltonian constraint. In fact, it may well be that given the
way in which $\beta$ enters eq.~(\ref{cix}), the Wheeler De Witt
equation $\hat{\cal H}\cdot\Psi =0$ will turn out to have solutions
only for particular values of $\beta$, a possibility that could be
investigated in Thiemann's proposed realisation of $\hat{\cal
H}$\cite{Tb}.

That these two ways may be related, or be the same, is indicated by
another remarkable Poisson bracket identity deduced by Thiemann:
\begin{equation} 
\left\{ {\epsilon_{ijk}\over\sqrt{\det E}}
E^{ia}E^{jb} F^{(\beta )k}_{ab}, \sqrt{\det E} \right\}=
2\beta^2\kappa E^{ia}K^i_a\delta (x,y) 
\end{equation} 
If at the quantum level we could handle the Hamiltonian
constraint and the volume, and translate this equation, we would 
be able to implement the "Wick rotation" 
from the Barbero to the Ashtekar connection representation.

At the same time, it is also important to establish whether the
discreteness of areas and volumes follows from the kinematics, which
embody the equivalence principle, or requires the Hamiltonian
constraint, i.e. the Einstein equations: because the discreteness of
areas may be used to establish the proportionality between area and
entropy\cite{Ro}, and this relation can be used to derive the Einstein
equations themselves, understood as equations of state\cite{Ja}.
\section{Conclusion} 
We have seen that one can formulate canonical
gravity as either a $SL(2,C)$ or an $SU(2)$ connection theory, but
that in the latter case an arbitrary parameter $\beta$ occurs in the
basic Poisson brackets; at the same time, present mathematical
technics can only cope with a quantum theory based on the group
$SU(2)$. However, no meaningful result on the spectra of operators can
be obtained unless we either fix this parameter, or learn how to
handle a theory based on the group $SL(2,C)$.
\section*{Acknowledgements} 
This work is part of a talk given at the
II Conference on Constrained Systems and Quantum Gravity, which took
place in Santa Margherita Ligure last september. I would like to thank
Prof. Vittorio De Alfaro for inviting me to talk at this meeting, and
Carlo Rovelli for comments and discussions.

\end{document}